\documentclass[12pt,overcite]{article}
\usepackage{txfonts}
\usepackage{txfonts}
\usepackage{txfonts}
\usepackage{txfonts}
\usepackage{txfonts}
\usepackage{txfonts}
\usepackage{txfonts}
\usepackage{txfonts}
\usepackage{txfonts}
\usepackage{txfonts}
\usepackage{txfonts}
\usepackage{txfonts}
\usepackage{txfonts}
\usepackage{txfonts}
\usepackage{txfonts}
\usepackage{txfonts}
\usepackage{mathrsfs}
\usepackage{amssymb}
\usepackage{amsfonts}
\usepackage{txfonts}
\usepackage[dvips]{graphicx}
\usepackage{epsf}
\usepackage{verbatim}
\begin{document}
\baselineskip=19pt
\begin{center}
{\bf\large Variable Selection Incorporating Prior Constraint
Information into Lasso}
\end{center}
\begin{center}
Shurong Zheng, Guodong Song and Ning-Zhong Shi\\
School of Mathematics $\&$ Statistics, Northeast Normal University,
P.R.China
\end{center}
\begin{center}
{\bf Abstract}
\end{center}
\par
We propose the variable selection procedure incorporating prior
constraint information into lasso. The proposed procedure combines
the sample and prior information, and selects significant variables
for responses in a narrower region where the true parameters lie. It
increases the efficiency to choose the true model correctly. The
proposed procedure can be executed by many constrained quadratic
programming methods and the initial estimator can be found by least
square or Monte Carlo method. The proposed procedure also enjoys
good theoretical properties. Moreover, the proposed procedure is not
only used for linear models but also can be used for generalized
linear models({\sl GLM}), Cox models, quantile regression models and
many others with the help of Wang and Leng (2007)'s LSA, which
changes these models as the approximation of linear models. The idea
of combining sample and prior constraint information can be also
used for other modified lasso procedures. Some examples are used for
illustration of the idea of incorporating prior constraint
information in variable selection procedures.

\par\noindent
{\bf 1. Introduction}
\par
In practice, a number of variables are included into an initial
regression analysis, but many of them may not be significant to the
response variables and should be excluded from the final model in
order to increase the accuracy of prediction and interpretation.
Variable selection is fundamental in statistical modeling. The least
absolute shrinkage and selection operator ({\sl LASSO}) (Tibshirani
\newline
{\footnotesize \qquad { {\sl Address for correspondence}: Ning-Zhong
Shi, School of Mathematics $\&$ Statistics,
 Northeast Normal
University, Changchun City 130024, P.R.China\\
\qquad Email: shinz@nenu.edu.cn}}
\newpage
1996) is a useful and well-studied approach to the problem of
variable selection (Knight and Fu 2000; Fan and Li 2001; Leng et al.
2006; Wang et al. 2007a; Yuan and Lin 2007). It shrinks some
coefficients and sets others to 0, and hence tries to retain the
good features of both subset selection and ridge regression.
Moreover, lasso's major advantage is its simultaneous execution of
both parameter estimation and variable selection. In particular,
allowing an adaptive amount of shrinkage for each regression
coefficient results in an estimator which is as efficient as oracle
(Zou 2006; Wang et al. 2007b; Wang and Leng 2007). About the
computational techniques, please see Osborne et al. (2000), Efron et
al. (2004), Rosset (2004), Zhao and Yu (2004) and Park and Hastie
(2006).
\par
In spite of that, in variable selection or the estimation of
regression coefficients, except for sample information, some prior
constraint information can be known. Constraints can be expressed as
 ${\bf g}(\beta)\leq 0$ including equalities and inequalities where ${\bf g}(\cdot)$  are
k-dimensional linear or nonlinear functions (see Rao and Toutenburg
1995; Silvapulle and Sen 2005). In fact, the common simple order
$\beta_1\leq\cdots\leq\beta_p$; tree order $\beta_i\leq\beta_p$ for
$i=1,\cdots,p-1$; umbrella order
$\beta_1\leq\cdots\leq\beta_l\geq\cdots\geq\beta_p$ or more
generally $A\beta\leq {\bf a}$ are only the special cases of ${\bf
g}(\beta)\leq 0$. All these constraints have very important
applications in biomedical studies, life science, econometrics and
social research etc. For example, in many biomedical studies,
treatment groups in a clinical trial many be formulated according to
increasing levels of dosage of a drug and the severity of disease in
patients. In econometrics, the homogeneity of degree zero of a
demand equation implies that the price and income elasticities add
up to zero, whereas the negativity of the substitution matrix in
consumer demand theory requires that all latent roots of the
substitution matrix should be nonpositive. Stahlecker (1987) shows a
variety of examples from the field of economics (such as
input-output models), where the constraints for the parameters are
so-called workability conditions of the form $\beta_i\geq 0$ or
$\beta_i\in (a_i,b_i)$ or $E(y_t|X)\leq a_t$. Literature deals with
this problem under the generic term constrained least squares (see
Judge and Takayama 1966; Dufour 1989; Geweke 1986; Moors and van
Houwelingen 1987; Rao and Toutenburg P75 1995). Dorfman and McIntosh
(2001) show that imposing the curvature conditions on a system of
demand equations improves the MSEs on estimated elasticities from 2
to $50\%$ depending on the signal-to-noise ratio and the sample
size. For researchers, it will increase the efficiency of variable
selection and parameter estimation to effectively combine the sample
and prior information because prior information tells us a narrower
region to select these variables and estimate these parameters.
\par
This paper proposes a procedure to combine prior and sample
information into lasso and hopes to obtain more accurate variable
selection and parameter estimation. The idea of combining prior
constraint and sample information can be shown by the black region
in Figure 1.
%\begin{figure}
%\centerline{\includegraphics[width=0.8\textwidth,
%angle=0]{graph1.EPS}}
%\begin{center}
%Figure 1
%\end{center}
%\end{figure}
It shows that when we know some prior information of parameters,
then variable selection will be executed in a narrower black region
AEFD not in a wide region ABCD. It will increase the efficiency of
choosing the true model correctly. Moreover, our procedure
incorporating prior constraint information is not only used for
linear models but also can be used for generalized linear models,
Cox models and quantile regression models with the help of Wang and
Leng (2007)'s LSA, which changes these models as the approximation
of linear models. In fact, prior constraint information can be also
used for other modified lasso procedures, e.g. Tibrashini et al.
(2005)'s fused lasso and the modified lasso procedure for an
adaptive amount of shrinkage for each regression coefficient (Zou
2006; Wang et al. 2007b; Wang and Leng 2007) etc.
\par
The paper is organized as follows: Section 2 introduces variable
selection procedure combining sample and prior constraint
information in lasso and other modified lasso procedures. Main
theoretical properties are discussed in Section 3. Section 4
discusses degrees of freedom of the lasso procedure incorporating
prior constraint information. The proposed procedure is illustrated
by some examples in Section 5. Section 6 gives a short discussion.

%\vskip 0.5cm\par\noindent {\bf 2.\quad Fused Lasso under Linear
%Inequality constraints}
%\par
%The following Figure 1 gives a schematic view of the fused lasso
%under linear inequality constraints.

\begin{flushleft}
{\bf 2.\quad Variable Selection Combining Sample and Prior
Constraint Information into Lasso}
\end{flushleft}
\par\noindent
{\bf 2.1\quad Variable Selection Combining Sample and Prior
Constraint Information into Lasso in Linear Models}
\par
We first consider variable selection incorporating prior constraint
information into lasso in linear models:
$$
\min\limits_{\beta}\sum\limits_{i=1}^{n}\left(y_i-{\bf
x}_i^{T}\beta\right)^2\quad\mbox{subject
to}\quad\sum\limits_{j}|\beta_j|\leq s\quad\mbox{and}\quad {\bf
g}(\beta)\leq {\bf 0}
$$
or
\begin{equation}
\hat{\beta}_s=Arg\left\{\min\limits_{\beta}\left({\bf Y}-{\bf
X}^{T}\beta\right)^{T}\left({\bf Y}-{\bf
X}^{T}\beta\right)\quad\mbox{subject
to}\quad\sum\limits_{j}|\beta_j|\leq s\quad\mbox{and}\quad {\bf
g}(\beta)\leq {\bf 0}\right\}.\label{b1}
\end{equation}
where ${\bf Y}=(y_1,\cdots,y_n)^{T}$, ${\bf X}=({\bf
x}^{T}_1,\cdots,{\bf x}^{T}_n)^{T}$ and ${\bf g}(\cdot)$ are linear
or nonlinear functions. That is, the modified lasso objective
function is as follows
$$
\min\limits_{\beta}\sum\limits_{i=1}^{n}(y_i-{\bf
x}_i^{T}\beta)^2+\lambda^{(1)}\sum\limits_{j}|\beta_j|+(\lambda^{(2)})^{T}{\bf
g}(\beta)
$$
where $\lambda^{(1)}$ and
$\lambda^{(2)}=(\lambda^{(2)}_1,\cdots,\lambda^{(2)}_k)^{T}$ are
tuning parameters. The tuning parameters can be obtained by
estimating the prediction error for the procedure incorporating
prior constraint information into lasso by cross-validation ({\sl
CV}) as described in chapter 17 of Efron and Tibshirani (1993) or
generalized cross-validation ({\sl GCV}). The prediction error of
prediction term $\hat{\eta}({\bf X})$ of CV is given by
$$
PE=E\{Y-\hat{\eta}({\bf X})\}^2.
$$
Then the value $\hat{s}$ yielding the lowest estimated PE is
selected.
\par
 In the following, we introduce how to choose the tuning parameters from CV in detail.
 Similarly, GCV can be used to choose the tuning parameters.
 $l$-fold CV is one of the methods
 to choose the tuning parameters $s$. $l$-fold
 CV is to split the n patterns into a
training set of size $n-l$ and a test of size $l$. $l$-fold CV
averages the squared error on the left-out pattern over all the
possible ways of obtaining such a partition. The advantage is that
all the data can be used for training - none has to be held back in
a separate test set. Take $l=1$ for an example. Let
\begin{equation}
\widehat{\beta}^{(-j)}_{s}=Arg\left\{\min\limits_{\beta}\sum\limits_{i=1,i\not=j}^{n}(y_i-{\bf
x}_i^{T}\beta)^2\quad\mbox{subject to}
\quad\sum\limits_{h}|\beta_h|\leq s\quad\mbox{and}\quad {\bf
g}(\beta)\leq {\bf 0}\right\}\label{b2}
\end{equation}
where $\widehat{\beta}^{(-j)}_{s}$ is the estimation on the training
data ${\bf x}_1$, $\cdots$, ${\bf x}_{j-1}$, ${\bf x}_{j+1}$,
$\cdots$, ${\bf x}_n$ for $j=1,\cdots,n$ from the procedure
incorporating prior constraint information into lasso. Let
$PE_s=\sum\limits_{j=1}^{n}\left(y_j-{\bf
x}_j^{T}\widehat{\beta}^{(-j)}_{s}\right)^2$ be the estimated
prediction error of $1$-fold CV given the tuning parameter $s$. Then
the chosen tuning parameters $s$ is as follows
$$
\hat{s}=Arg\left\{\min\limits_{s}PE_s\right\}=Arg\left\{\min\limits_{s}\sum\limits_{j=1}^{n}\left(y_j-{\bf
x}_j^{T}\widehat{\beta}^{(-j)}_{s}\right)^2\right\}
$$
where $\hat{s}$ minimizes the estimated prediction error. Then the
simultaneous parameter estimation and variable selection
incorporating prior constraint information is as follows
\begin{equation}
\widehat{\beta}_{\hat{s}}=Arg\left\{\min\limits_{\beta}\sum\limits_{i=1}^{n}(y_i-{\bf
x}_i^{T}\beta)^2\quad\mbox{subject
to}\quad\sum\limits_{h}|\beta_h|\leq \hat{s}~~\mbox{and}~~ {\bf
g}(\beta)\leq {\bf 0}\right\}.\label{b3}
\end{equation}
\par
{\sl\bf Remark} 1. ({\sl Algorithm})We know that the most important
thing for obtaining $\widehat{\beta}_{\hat{s}}$ is to compute
$\widehat{\beta}^{(-j)}_{s}$. If there are no constraints on the
parameters, many well developed procedures can be used to find the
solution for
$$
\min\limits_{\beta}\sum\limits_{i=1,i\not=j}^{n}(y_i-{\bf
x}_i^{T}\beta)^2\quad\mbox{subject to}
\quad\sum\limits_{h}|\beta_h|\leq s.$$ For example, quadratic
programming (Tibshirani 1996), the shooting algorithm (Fu 1998),
local quadratic approximation (Fan and Li 2001) and lease angle
regression ({\sl LARS}) (Efron et al. 2004). When there are prior
constraint information, the above procedures can not be directly
used for $(\ref{b2})$. But if some modifications are made for these
algorithms, $(\ref{b2})$ may be solved by them. It will be an
interesting topic for us in the future. In fact, many quadratic
programming methods can be used to find the solution for
$(\ref{b2})$ (see Dantig and Eaves 1974). The solution of the
quadratic programming does not yield a sparse solution. If a
tolerance is set, the small parameter estimate can be regarded as 0.
\par
{\sl\bf Remark} 2. ({\sl Initial Estimator}) In fact, the OLS
estimator may be regarded as the initial estimator. But in order to
obtain more accurate estimator, Monte Carlo method can be used for
the initial estimator of $(\ref{b1})$ or $(\ref{b2})$. The optimal
problem $(\ref{b1})$ can be written as
$$
\begin{array}{ll}
({\bf Y}-{\bf X}\beta)^{T}({\bf Y}-{\bf X}\beta)
&=\beta^{T}{\bf X}^{T}{\bf X}\beta-2\beta^{T}{\bf X}^{T}{\bf Y}+{\bf Y}^{T}{\bf Y}\\
\\
&=(\beta-\mu)^{T}\Sigma^{-1}(\beta-\mu)+{\bf Y}^{T}\left(I-{\bf
X}({\bf X}^{T}{\bf X})^{-1}{\bf X}^{T}\right){\bf Y}
\end{array}
$$
with known $\mu_{p\times 1}=\left({\bf X}^{T}{\bf X}\right)^{-1}{\bf
X}^{T}{\bf Y}$ and $\Sigma_{p\times p}=\left({\bf X}^{T}{\bf
X}\right)^{-1}$. That is,
$$
\widehat{\beta}_{s}=Arg\left\{\min\limits_{\beta}(\beta-\mu)^{T}\Sigma^{-1}(\beta-\mu)
\quad\mbox{subject~to}\quad \sum\limits_{h=1}^{p}|\beta_h|\leq s
~\mbox{and}~{\bf g}(\beta)\leq{\bf 0}\right\}
$$
or
\begin{equation}
\widehat{\beta}_{s}=Arg\max\limits_{\beta}-\frac{1}{2}
\left[(\beta-\mu)^{T}\Sigma^{-1}(\beta-\mu)+\log(|\Sigma|)\right]
\quad\mbox{subject~to}\quad \left\{
\begin{array}{l}
\sum\limits_{j=1}^{p}|\beta_j|\leq s\\
{\bf g}(\beta)\leq{\bf 0}
\end{array}
\right. \label{b4}
\end{equation}
where
\begin{equation}
l(\beta)=-\frac{1}{2}(\beta-\mu)^{T}\Sigma^{-1}(\beta-\mu)-\frac{1}{2}\log(|\Sigma|)\label{b45}
\end{equation}
is just the log-density of $N(\mu, \Sigma)$ regarding $\beta$ as a
random variable. Randomly draw $m=100000$ samples ${\bf
Z}_1,\cdots,{\bf Z}_m$ from $N(\mu,\Sigma)$ where ${\bf
Z}_j=(Z_{1j}$, $\cdots$, $Z_{pj})^{T}$ for $j=1,\cdots,m$. Set
$Z_{old}$ as the initial estimator which satisfies
$$
Z_{old}=Arg\left\{\max\limits_{j=1,\cdots,m}l(Z_j)\quad\mbox{subject~to}\quad
\sum\limits_{h=1}^{p}|Z_{hj}|\leq s ~\mbox{and}~{\bf g}(Z_j)\leq{\bf
0}\right\}.
$$

\par\noindent
{\bf 2.2\quad Variable Selection Combining Sample and Prior
Constraint Information into Other Modified Lassos}
\par
 The limitation of lasso is that
 all the regression coefficients share the same amount of shrinkage
$\min\limits_{\beta}\sum\limits_{i=1}^{n}\left(y_i-{\bf
x}_i^{T}\beta\right)^2+\lambda\sum\limits_{j=1}^{p}|\beta_j|.$ Then
Wang et al. (2007b) extend the lasso to the modified lasso$^*$
criterion
 which allows for different tuning parameters for different
 coefficients
$$
\min\limits_{\beta}\sum\limits_{i=1}^{n}\left(y_i-{\bf
x}_i^{T}\beta\right)^2+\sum\limits_{j=1}^{p}\lambda_j|\beta_j|.
$$
In order to combining the sample and prior constraint information,
variable selection procedure can be executed as follows
$$
\min\limits_{\beta}\sum\limits_{i=1}^{n}\left(y_i-{\bf
x}_i^{T}\beta\right)^2+\sum\limits_{j=1}^{p}\lambda_j|\beta_j|+\phi^{T}{\bf
g}(\beta)
$$
which not only uses the prior information but also overcomes the
limitation of the traditional lasso procedure.
\par
Similarly, the prior constraint information can be incorporated into
Tibshirani et al. (2005)'s fused lasso which encourages sparsity in
their differences, i.e. flatness of the coefficient profiles
$\beta_j$ as a function of $j$.
\par\noindent
{\bf 2.3\quad Variable Selection Combining Sample and Prior
Constraint Information into Lasso in Nonlinear Models}
\par
The proposed variable selection procedure can not be directly used
for nonlinear models, e.g. generalized linear models; Cox models and
quantile regression models etc. But with the help of Wang and Leng
(2007)'s {\sl LSA}, the proposed variable selection procedure can be
used for these nonlinear models. LSA regards
\begin{equation}
(\beta-\tilde{\beta})^{T}\hat{\Sigma}^{-1}(\beta-\tilde{\beta})\label{b5}
\end{equation}
as the least square approximation of the original loss
$n^{-1}L_n(\beta)$ where $\tilde{\beta}$ is the unpenalized
estimator obtained by minimizing $L_n(\beta)$,
$\hat{\Sigma}^{-1}=n^{-1}\ddot{L}_n(\tilde{\beta})$ and
$\ddot{L}_n(\cdot)$ is the second derivatives of the loss function
$L_n(\cdot)$. The expression $(\ref{b5})$ is similar to the
log-density $l(\beta)$ in $(\ref{b45})$. So it is clear that the
lasso procedure incorporating prior constraint information can also
be used for variable selection in nonlinear models with the help of
the least squares approximation.
\begin{flushleft}
{\bf 3.\quad Some Theoretical Properties}
\end{flushleft}
\par
In this section, we derive some theoretical results for the lasso
combining the sample and prior constraint information that are
analogous to those for the lasso and fused lasso (Knight and Fu
(2000); Tibshirani et al (2005)). The penalized least squares
criterion is
$$
\sum\limits_{i=1}^{n}(y_i-X_i^{T}\beta)^2+\lambda_n^{(1)}\left(\sum\limits_{j=1}^{p}|\beta_j|-s\right)+
{\bf g}(\beta)^{T}\lambda_{n}^{(2)}
$$
with $\beta=(\beta_1,\cdots,\beta_p)^{T}$ and ${\bf
x}_i=(x_{i1},\cdots,x_{ip})^{T}$, and the Lagrange multipliers
$\lambda_n^{(1)}$ and $\lambda_n^{(2)}$ are functions of the sample
size $n$. Let the optimal solution be $\widehat{\beta}_n$.
\par
For simplicity, we assume that $p$ is fixed as $n\rightarrow \infty$
and ${\bf g}(\cdot)$ are differential convex functions. The
following theorem adequately illustrates the basic dynamics of the
lasso combining sample and prior constraint information.
\par
 {\bf Theorem 1}.\quad If
$\lambda_{n}^{(l)}/\sqrt{n}\rightarrow\lambda_0^{(l)} (l=1,2)$ and
$$
C=\lim\limits_{n\rightarrow\infty}\left(\frac{1}{n}\sum\limits_{i=1}^{n}X_{i}X_i^{T}
\right)
$$
is non-singular, then
$$
\sqrt{n}(\widehat{\beta}_n-\beta)\stackrel{D}{\rightarrow}\arg
\min\limits_{u}V(u)
$$
where $$
V(u)=-2u^{T}W+u^{T}Cu+\lambda_0^{(1)}\sum\limits_{j=1}^{p}\{u_jsgn(\beta_j)
I(\beta_j\not=0)\}+|u_j|I(\beta_j=0)+\left(\frac{\partial {\bf
g}(\beta)}{\partial \beta}u\right)^{T}\lambda_0^{(2)}$$ and ${\bf
W}$ has an $n(0,\sigma^2C)$ distribution.
\par
Proof.
$$
\sum\limits_{i=1}^{n}(y_i-X_i^{T}\beta)^2+\lambda_n^{(1)}\left(\sum\limits_{j=1}^{p}|\beta_j|-s\right)+
{\bf g}(\beta)^{T}\lambda_{n}^{(2)}
$$
where $\lambda_{n}^{(1)}$ and $\lambda_{n}^{(2)}$ are functions of
the sample size $n$. Define $V_n(u)$ by
$$
V_n(u)=\sum\limits_{i=1}^{n}\{(\varepsilon_i-u^{T}x_i/\sqrt{n})^2-\varepsilon_i^2\}+\lambda_n^{(1)}
\sum\limits_{j=1}^{p}\left(|\beta_j+u_j/\sqrt{n}|-|\beta_j|\right)+
\left({\bf g}(\beta+u/\sqrt{n})-{\bf
g}(\beta)\right)^{T}\lambda_n^{(2)}
$$
with $u=(u_1,\cdots,u_p)^{T}$ and note that $V_n(u)$ is minimized at
$\sqrt{n}(\widehat{\beta}_n-\beta)$. First note that
$$
\sum\limits_{i=1}^{n}\{(\varepsilon_i-u^{T}x_i/\sqrt{n})^2-\varepsilon_i^2\}\stackrel{D}{\rightarrow}
-2u^{T}W+u^{T}Cu
$$
with finite dimensional convergence holding trivially where
$$C=\lim\limits_{n\rightarrow\infty}\left(\frac{1}{n}\sum\limits_{i=1}^{n}X_{i}X_i^{T}
\right)\quad\mbox{and}\quad W\sim n(0,\sigma^2C).$$ We also have
$$
\lambda_n^{(1)}\sum\limits_{j=1}^{p}
\left(|\beta_j+u_j/\sqrt{n}|-|\beta_j|\right)\stackrel{D}{\rightarrow}
\lambda_0^{(1)}\sum\limits_{j=1}^{p}\{u_jsgn(\beta_j)
I(\beta_j\not=0)\}+|u_j|I(\beta_j=0)
$$
and
$$
\left({\bf g}(\beta+u/\sqrt{n})-{\bf
g}(\beta)\right)^{T}\lambda_n^{(2)}=\left(\frac{\partial {\bf
g}(\beta)}{\partial \beta}u\right)^{T}\lambda_0^{(2)}.
$$
Thus $V_n(u)\stackrel{D}{\rightarrow}V(u)$, with finite dimensional
convergence holding trivially where
$$
V(u)=-2u^{T}W+u^{T}Cu+\lambda_0^{(1)}\sum\limits_{j=1}^{p}\left[u_jsgn(\beta_j)
I(\beta_j\not=0)\right]+|u_j|I(\beta_j=0)+\left(\frac{\partial {\bf
g}(\beta)}{\partial \beta}u\right)^{T}\lambda_0^{(2)}.$$ Since $V_n$
is convex and $V$ has a unique minimum, it follows (Geyer, 1996)
that
$$
\arg\min\limits_{u}V_n(u)=\sqrt{n}(\widehat{\beta}_n-\beta)\stackrel{D}{\rightarrow}\arg
\min\limits_{u}V(u).
$$
\par
{\bf Theorem 2.}\quad The procedure incorporating prior constraint
information into lasso will increase efficiency of selecting
significant variables for responses compared with the traditional
lasso procedures.
\par
Proof. Theoretically, the general lasso procedure is as follows
$$
\tilde{\beta}_{\tilde{s}}=Arg\left\{\min\limits_{\beta}\sum\limits_{i=1}^{n}(y_i-{\bf
x}_i^{T}\beta)^2\quad\mbox{subject
to}\quad\sum\limits_{h}|\beta_h|\leq \tilde{s}\right\}
$$
where GCV or CV is used to choose the tuning parameter $\tilde{s}$
which minimizes the estimated prediction errors
$$
\tilde{s}=Arg\left\{\min\limits_{s} PE_s\right\}.$$ If the estimator
$\tilde{\beta}_{\tilde{s}}$ satisfies prior constraints $ {\bf
g}(\beta_{\tilde{s}})\leq {\bf 0}$, it means that
$\tilde{\beta}_{\tilde{s}}$ clearly minimizes the estimated
prediction errors in a narrower region ${\bf g}(\beta)\leq {\bf 0}$.
That is,
$$
\tilde{\beta}_{\tilde{s}}=\widehat{\beta}_{\hat{s}}
$$
where $\widehat{\beta}_{\hat{s}}$ is the estimator of parameter by
the lasso procedure incorporating prior constraint information in
$(\ref{b3})$. Now, we take Figure 1 as an example. From Figure 1, we
know that $\tilde{\beta}_{\tilde{s}}$ lies in the region $ABCD$ and
minimizes the estimated prediction errors. Moreover, we know that
$\tilde{\beta}_{\tilde{s}}$ lies in the region $AEFD$. It is clear
that $\tilde{\beta}_{\tilde{s}}$ minimizes the estimated prediction
errors in the region above the line $EF$. Furthermore, the true
model also lies in the region above the line $EF$. So we obtain that
if $\tilde{\beta}_{\tilde{s}}$ selects the true variables correctly,
that is, the nonzero components of $\tilde{\beta}_{\tilde{s}}$ are
just the significant covariates, then $\widehat{\beta}_{\hat{s}}$
also selects the true variables correctly.
\par
If $\tilde{\beta}_{\tilde{s}}$ doesn't select significant variables
correctly, some prior constraint information may bring us into a
narrower region to select these variables again. It will increase
the efficiency of variable selection.

\begin{flushleft}
{\bf 4.\quad Standard error and degrees of freedom of the lasso
estimate}
\end{flushleft}
\par
Since our lasso procedure combining sample and prior constraint
information is a nonlinear and nondifferentiable function of the
response values even for a fixed value of $s$, it is difficult to
obtain an accurate estimate of its standard error. The problem can
be solved by bootstrap approach: either $s$ can be fixed or we may
optimize over $s$ for each bootstrap sample.
\par
Efron et al. (2004) consider a definition of degrees of freedom
using the formula of Stein (1981):
$$
df(h(y))=\frac{1}{\sigma^2}\sum\limits_{i=1}^{n}cov(y_i,h_i)=E\left\{
\sum\limits_{i=1}^{n}\frac{\partial h(y)}{\partial y_i}\right\}
$$
where $y=(y_1,\cdots,y_n)^{T}$ is a multivariate normal vector with
mean $\mu$ and covariance I, and $h(y)$ is an estimator, an almost
differential function from $\mathbb{R}^{n}$ to $\mathbb{R}^{n}$. For
the lasso with orthonormal design ${\bf X}^{T}{\bf X}={\bf
I}_{p\times p}$, the degrees of freedom are the number of non-zero
coefficients. Tibshirani et al.(2005) show that the natural estimate
of the degrees of freedom of the fused lasso is
$$
\begin{array}{ll}
df(\hat(y))&=\#\{\mbox{non-zero coefficient block in} \hat{\beta}\}\\
\\
           &=p-\#\{\beta_j=0\}-\#\{\beta_j-\beta_{j-1}=0, \beta_j, \beta_{j-1}\not=0\}
\end{array}
$$
similarly, the natural estimate of the degrees of freedom of the
lasso incorporating prior constraint information is
$$
df(\hat{y})=p-\#\{\beta_j=0\}-\#\{{\bf g}(\beta)=0\}.
$$
The degrees can be used for BIC-type tuning parameter selector.

\begin{flushleft}
{\bf 5.\quad Some Examples }
\end{flushleft}
\par
In the following, we give three examples for illustration of the
proposed procedure's practical applications in many models.
\par\noindent
Example 1: {\bf linear inequality constraints in linear models}
\par
Wolak (1989) or (Silvapulle and Sen 2005 P9) consider the following
double-log demand function
$$
\log Q_t=\alpha+\beta_1\log PE_t+\beta_2 \log PG_t+\beta_3 \log
I_t+\gamma_1 D1_t+\gamma_2 D2_t+\gamma_3 D3_t+\epsilon_t
$$
which is a linear model where
$$
\begin{array}{l}
~~Q_t=\mbox{aggregate electricity demand},\\
\\
PE_t=\mbox{average price of electricity to the residential sector},\\
\\
PG_t=\mbox{price of natural gas to the residential sector},\\
\\
~~I_t=\mbox{income per capita},\\
\\
\mbox{and}~ D1_t, D2_t, D3_t~ \mbox{are seasonal dummy variables}.
\end{array}
$$
Prior knowledge suggests that
$$
\left(
\begin{array}{lll}
1&0&0\\
0&1&0\\
0&0&1
\end{array}
\right) \left(
\begin{array}{l}
\beta_1\\
\beta_2\\
\beta_3
\end{array}\right)\geq
\left(
\begin{array}{l}
0\\
0\\
0
\end{array}\right)$$ which are linear inequality constraints. A typical model
selection question is whether or not the foregoing model provides a
better fit than the simpler model
$$
\log Q_t=\alpha+\gamma_1 D1_t+\gamma_2 D2_t+\gamma_3
D3_t+\epsilon_t.
$$
Wolak (1989) or Wang et al. (2007b) discuss the model selection
problem by a test method or by a variable selection method,
respectively.
\par\noindent
Example 2: {\bf nonlinear inequality constraints in linear models}
\par
Dufour (1989) considers the following econometric model
$$
y_i=f(x_i,\beta)+\epsilon_i
=\beta_1+\beta_2x_{i2}+\beta_3x_{i3}+\beta_4x_{i2}^2+\beta_5x_{i3}^2+2\beta_6x_{i2}x_{i3}+\epsilon_i.
$$
This could be a production function or a unit cost function where
$y_i$ is the production or unit cost and $\{x_{i2},x_{i3}\}$ are
inputs. A problem of interest in econometrics is whether
$f(x_i,\beta)$ is concave in $x_i$, which can be expressed by the
following nonlinear inequality constraints
$$
\beta_4\leq 0,~ \beta_5\leq 0,~ \beta_4\beta_5-\beta_6^2\geq 0.
$$
Dufour (1989) discusses the model selection problem by a test
method.
\par\noindent
Example 3: {\bf linear equality and inequality constraints in
generalized linear models}
\par
An assay was carried out with the bacterium {\sl E. coli} strain
343/358(+) to evaluate the genotoxic effects of 9-aminoacridine
(9-AA) and potassium chromate (KCr). Piegorsch (1990) and Silvapulle
(1994) consider the following log-linear model
\begin{equation}
\log (1-\pi_{ij})=\mu+\alpha_i+\tau_j+\eta_{ij}.\label{b8}
\end{equation}
to evaluate whether potassium chromate and 9-AA have a synergistic
effect where $i=1,2$, $j=1,\cdots,5$ and
$$\pi_{ij}=Pr\{\mbox{positive response for a test unit in cell
(i,j)}\}.$$ In fact, the log-linear model is just logistic
regression model which is one of generalized linear models({\sl
GLM}). To ensure that the parameters in $(\ref{b8})$ are identified,
Piegorsch (1990) and Silvapulle (1994) impose the constraints
$\alpha_1=\tau_1=\eta_{i1}=\eta_{1j}=0$ for all $(i,j)$ and
$$\left(
\begin{array}{llll}
1&0&0&0\\
0&1&0&0\\
0&0&1&0\\
0&0&0&1
\end{array}
\right) \left(
\begin{array}{l}
\eta_{22}\\
\eta_{23}\\
\eta_{24}\\
\eta_{25}
\end{array}
\right) \geq \left(
\begin{array}{l}
0\\
0\\
0\\
0
\end{array}\right)$$ which means that potassium chromate and 9-AA have a
synergistic effect. The model selection problem is analyzed by a
test in Piegorsch (1990), Silvapulle (1994) and Silvapulle and Sen
(2005 P161).

\begin{center}
{\bf 6.\quad Discussion}
\end{center}
\par
We proposed a modified lasso procedure combining prior constraint
and sample information for variable selection and parameter
estimation. The proposed procedure increases the efficiency of
choosing the true model correctly because it executes variable
selection and parameter estimation in a narrower region where the
true parameters lie. The procedure may be computed by many quadratic
programming methods.
\par
Moreover, the idea of incorporating prior constraint information can
be used for other lasso procedures, e.g. fused lasso and modified
lasso procedure for an adaptive amount of shrinkage for each
regression coefficient.
\par
More work remains to be done. Efron et al. (2004)'s LARS is a good
computational procedure which only needs $p$ steps. But now it is
not directly used for the lasso procedure incorporating prior
constraint information. In our procedure, Monte Carlo estimator can
be used for the initial estimator. How to extend LARS to the lasso
procedure incorporating prior constraint information is an
interesting topic for future study.

\begin{center}
{\bf References}
\end{center}

\par\noindent
Dantig, G. B. and Eaves, B. C.(1974) {\sl Studies in Optimization.}
Washington: Mathematical Association of America.

\par\noindent
Dorfman, J. and McIntosh, C.S. (2001) Imposing inequality
restrictions: efficiency gains from economic theory. {\sl Economics
Letters}, {\bf 71}(2), 205-209.

\par\noindent
Dufour, J.M. (1989) Nonlinear hypotheses, inequality restrictions
and non-nested hypotheses: Exact simultaneous tests in linear
regression. {\sl Econometrics}, {\bf 57}, 335-355.

\par\noindent
Efron, B., Hastie, T., Johnstone, I. and Tibshirani, R. (2004) Least
angle regression. {\sl The Annals of Statisitcs}, {\bf 32}, 407-489.

\par\noindent
Efron, B. and Tibshirani, R. (1993) {\sl An introduction to the
bootstrap}. London: Chapman and Hall.

\par\noindent Fan, J. and Li, R. (2001) Variable selection via nonconcave
penalized likelihood and its oracle properties. {\sl Journal of the
American Statistical Association}, {\bf 96}, 1348-1360.

\par\noindent
Fu, W.J. (1998) Penalized regression: the bridge versus the lasso.
{\sl J. Comput. and Graph. Statist.}, {\bf 7}, 397-416.

\par\noindent
Geyer, C. (1996) On the asymptotics of convex stochastic
optimization. {\sl Technical Report}. University of Minnesota,
Minneapolis.
\par\noindent
Geweke, J. (1986) Exact inference in the inequality constrained
normal linear regression model. {\sl J. Applied Econometrics}, {\bf
1}, 127-141.
\par\noindent
Gill, P.E., Murray, W. and Saunders, M. A.(1997) Users guide for
SqOPT 5.3: a Fortran package for large-scale linear and quadratic
programming. {\sl Technical report NA 97-4.} University of
California, San Diego.

\par\noindent
Judge, G.G. and Takayama, T. (1966) Inequality restrictions in
regression analysis. {\sl J. Amer. Statist. Assoc.}, {\bf 61},
166-181

\par\noindent
Knight, K. and Fu, W. (2000) Asymptotics for lasso-type estimators.
{\sl Ann. Statist.} {\bf 28}, 1356-1378.

\par\noindent Leng, C., Lin, Y., and Wahba, G. (2006) A note on lasso and
related procedures in model selection. {\bf Statistica Sinica}, {\bf
16}, 1273-1284.

\par\noindent
Moors J.J.A. and van Houwelingen, J.C. (1987) Estimation of linear
models with inequality restrictions. Research Report FEW 291,
Tilburg University.

\par\noindent Osborne, M.R., Presnell, B. and Turlach, B.A. (2000), A new
approach to variable selection in least squares problems. {\sl IMA
Journal of Numerical Analysis}, {\bf 20}, 389-404.

\par\noindent
Park, M.Y. and Hastie, T. (2006) An L1 regularization-path algorithm
for generalized linear models. Manuscript, Department of Statistics,
Stanford University.

\par\noindent
Piegorsch, W.W. (1990) One-sided significance tests for generalized
linear models under dichotomous response. {\sl Biometrics}, {\bf
46}, 309-316.

\par\noindent
Rao C.R. and Toutenburg, H. (1995) {\sl Linear models}.
Springer-Verlag New York: Berlin Heidelberg

\par\noindent Rosset, S. (2004) {\sl Tracking curved regularized optimization
solution paths}. NIPS 2004.

\par\noindent
Silvapulle, M.J. (1994) On tests against one-sided hypotheses in
some generalized linear models. {\sl Biometrics}, {\bf 50}, 853-858.

\par\noindent
Silvapulle, M.J. and Sen, P.K.(2005) {\sl Constrained statistical
inference}. John Wiley $\&$ Sons, Inc., Hoboken: New Jersey.

\par\noindent
Stahlecker, P. (1987) A prior Information und
Minimax-Sch$\ddot{a}$tzung im linearen regressionsmodell.
Athen$\ddot{a}$um, Frankfurt/M.

\par\noindent
Stein, C. (1981) Estimation of the mean of a multivariate normal
distribution. {\sl Ann. Statist.}, {\bf 9}, 1131-1151.
\par\noindent
Tibshirani, R. (1996) Regression shrinkage and selection via the
lasso. {\sl J.R. Statist. Soc. B.,} {\bf 58}, 267-288.

\par\noindent
Tibshirani, R., Saunders, M., Rosset, S., Zhu, J. and Knight, K.
(2005) Sparsity and smoothness via the fused lasso. {\sl Journal of
the Royal Statistical Society, Series B.}, {\bf 67}, 91-108.

\par\noindent Wang, H. and Leng, C. (2007) Unified lasso estimation via
least squares approximation. {\sl Journal of the American
Statistical Association}, to appear.

\par\noindent Wang, H., Li, G. and Jiang, G. (2007a)
On the consistency of SCAD tuning parameters selector. {\sl
Biometrka}, to appear.

%\parnoindent Wang, H., Li, G.D. and Tsai, C.L. (2007b)
%Robust regression shrinkage and consistent variable selection via
%the LAD-LASSO. {\sl Journal of Business and Economics Statistics},
%to appear.

\par\noindent Wang, H., Li, G.D. and Tsai, C.L. (2007b)
Regression coefficient and autoregressive order shrinkage and
selection via lasso. {\sl Journal of the Royal Statistical Society,
Series B.}, {\bf 69}, 63-78.

\par\noindent Wolak, F.A. (1989) Testing inequality constraints in linear
econometric models. {\sl Journal of Econometrics}, {\bf 41},
205-235.

\par\noindent
Yuan, M. and Lin, Y. (2007) On the nonnegative garrote estimator.
{\sl Journal of the Royal Statistical Society, Series B.}, To appear

\par\noindent
Zhao, P. and Yu, B. (2004) Boosted lasso. {\sl Technical Report
$\#678$,} Statistics, UC Berkeley.

\par\noindent Zou, H. (2006) The adaptive lasso and its oracle properties.
{\sl Journal of the American Statistical Association,} {\bf 101},
1418-1429.

%\end{enumerate}

\end{document}